\documentclass[preprint,letter]{aastex}
\shorttitle{Konus catalog of short GRBs}
\shortauthors{Mazets et al.}
\begin{document}
\title{Konus catalog of short GRBs}
\author{E.P.Mazets, R.L. Aptekar, D.D. Frederiks, S.V. Golenetskii,\\ V.N. Il'inskii and V.D. Palshin}
\affil{Ioffe Physico-Technical Institute, Russian Academy of Sciences}
\affil{St.Petersburg, 194021, Russia}
\author{T.L.Cline and P.S. Butterworth}
\affil{NASA Goddard Space Flight Center, Greenbelt, Maryland, 20771, USA}
\begin{abstract}
Observational data on the short GRBs obtained with the GGS-Wind Konus 
experiment in the period from 1994 to 2002 are presented. The catalog 
currently includes 130 events, detailing their appearance rate, time 
histories, and energy spectra. Evidence of an early X-ray and 
gamma-ray afterglow for some of the short GRBs is discussed.
The catalog is  available electronically at 
\url{http://www.ioffe.ru/LEA/shortGRBs/Catalog/}.
\end{abstract}

\section{INTRODUCTION}
The gamma-ray burst (GRB) duration has a bimodal distribution 
(Mazets et~al. 1981; Norris et~al. 1984; Hurley 1992; Kouveliotou et~al. 1993).
This indicates the existence of two distinct morphological classes of events, 
namely short-duration (\textless~2~s) bursts  and long-duration (\textgreater~2~s) 
bursts. Approximately 20 per cent  of the observed bursts are short. 
Their  energy spectra are usually harder than the spectra of long bursts 
(Kouveliotou et al. 1993).

The catalog contains data  on 130 short GRBs observed  with the Konus-Wind 
experiment on the Wind spacecraft in 1994--2002. The catalog presents  time 
histories, energy spectra, fluences, peak fluxes, spectral parameters, and 
hardness ratios. Most of hardness ratios reveal spectral variability.
The catalog is  available electronically at 
\url{http://www.ioffe.ru/LEA/shortGRBs/Catalog/}.

Searches for the optical and radio afterglow of short GRBs have been carried 
out in only a few cases. Four GRBs were localized with  high accuracy by 
the Interplanetary Network~(IPN) (Hurley et al. 2002). No optical and radio 
afterglow emission was detected for these four events. No X-ray 
counterparts have been detected  so far for the short bursts localized by 
Beppo-SAX (Gandolfi et al. 2000) or by HETE-2 (Lamb et al. 2002). The rapid 
follow-up observations have resulted in only upper limits on the brightness 
of the afterglows from these GRBs.

At the same time early X-ray and gamma-ray afterglows of short
bursts were detected in a bumber cases by the GRB detectors themselves, in time 
intervals from seconds to tens of seconds after the trigger. BATSE observations 
showed that such a weak afterglow exists for some of the short bursts, 
lasting several tens seconds (Burenin, 2000; Lazatti et~al. 2001; 
Connaughton, 2002). The afterglows of short GRBs were also detected by 
the Konus-Wind experiment. Afterglow emission in the energy range 
bellow 1~MeV is seen for about 10 per cent of events. These bursts 
are included in the catalog. A statistical analysis of burst sampling
reveals that the afterglow is a more common feature of short GRBs. These 
results will be discussed in more detail elsewhere.

\section{OBSERVATIONS}
The Konus-Wind instrument is a gamma-ray spectrometer consisting of two 
identical gamma-ray detectors S1 and S2 which  observe the southern and 
northern ecliptic hemispheres respectively in all-sky monitoring mode. 
Each detector contains a NaI(Tl) crystal, 5'' in diameter and 3'' in height, 
in a housing with an entrance window made of beryllium. The nominal energy 
range of gamma-ray measurements covers the interval from 12~keV up to 10~MeV.
The instrument operates in two main modes: background measurements and 
triggered burst detections. While in the ``background'' mode, the count rates 
are recorded in three energy windows G1~(12--45~keV), G2~(45--190~keV), and 
G3~(190--770~keV). The time resolution of such observations is 2.94~s. When 
the count rate in the G2 window exceeds the 6$\sigma$ trigger threshold the 
instrument is switched into the ``burst'' mode. The count rates in the three 
energy windows are recorded with time resolution from 2~ms up to 256~ms. 
These time histories, of total duration 230~s, also include 0.512~s of 
pre-trigger history. The spectral measurements are carried out in two 
energy intervals, 12--760~keV and 200~keV~--~10~MeV, with 64 spectra being 
recorded for each interval over the 63-channel quasilog energy scale. The 
first four spectra are measured with a fixed accumulation time of 64~ms in 
order to study short bursts. An adaptive system varies the accumulation 
time for next spectra from 256~ms to 8.192~s depending on the current count 
rate in the G2 window. A detailed description of the Konus-Wind instrument 
and its operational modes is given in Aptekar et~al.~(1995). The  gain of the 
Konus-Wind detectors has slowly decreased during the long period of operation. 
Instrumental control of the gain was exhausted in 1997. 
The low energy threshold grown to 20~keV for the S1 detector and 
to 17.5~keV for the S2 detector, from the original 12~keV. For convenience of comparison, 
the data for fluences and peak fluxes have been reduced to a 15~keV threshold. 
Results of the Konus-A experiments were used for the analysis of 
several short bursts in the absence of the Konus-Wind data. The spectrometric 
detectors of the Konus-A instruments are identical to the Konus-Wind 
detectors. The Konus-A experiments were carried out on board the 
low-orbiting spacecraft Kosmos-2326 (1995--1997) and Kosmos-2367 (1999--2001).

\section{DATA PROCESSING}
The process of gamma-ray photon spectral deconvolution from the count rate 
spectra is based on the Monte Carlo response function calculations and the 
results of the laboratory calibration described by Terekhov et~al.~(1998). 
The response functions of the Konus-Wind detectors were calculated in the 
energy range 5~keV~--~10~MeV for angles of incidence in the 0$^\circ$--90$^\circ$ 
range. The angles between the GRB direction and the detector's 
axis (GRB angles) have been determined using IPN or BATSE localizations. 
The ratio of the S1 and S2 detector count rates were used to estimate 
the GRB angle in those cases when GRBs were not localized. The 
dependence of the GRB angle on the S1/S2 ratio was derived from bursts with 
known localization. It allows the evaluation of the GRB angle for bursts with 
an unknown localization with an accuracy of 5$^\circ$--20$^\circ$. The deconvolution 
procedure (Terekhov et al. 1998) permits fits to the bursts spectra using 
different proposed models or their combinations. Most of the appropriate 
spectral models of GRBs were considered by Preece et~al.~(2000). 
For convenience, we follow their notation:
\begin{enumerate}
\item  The power law model (PL): 
$$f(E)= \mathrm{d}N(E)/\mathrm{d}E=A(E/100) E^\alpha$$
\item  The broken power law model (BPL):
$$f(E) = \left\{
\begin{array}{ll}
A(E/100)^{\alpha}, & \mbox{if } E \leq E_b\\
A(E_b/100)^{\alpha} (E/E_b)^{\beta}, & \mbox{if } E>E_b
\end{array} \right.$$
\item      The COMP model:
$$f(E)=A(E/100)\exp(-E/E_0)$$
\item      The GRB model:
$$f(E)=\left\{
\begin{array}{ll}
A(E/100)^\alpha \exp(-E/E_0), & \mbox{if } E \leq (\alpha - \beta) E_0\\
A[(\alpha-\beta) E_0/100]^{(\alpha-\beta)} \exp(\beta - \alpha)(E/100)^\beta, & \mbox{if } E > (\alpha - \beta) E_0
\end{array} \right.$$

where: $A$ is the intensity coefficient in photons s$^{-1}$cm$^{-2}$keV$^{-1}$,
       $E_b$  is the break energy in keV,
       $E_0$ is the energy parameter in keV
       and $\alpha$, $\beta$ are the spectral indexes,
       $E$ is the photon energy in keV.
\end{enumerate}

\section{CATALOG}
\subsection{Catalog Structure}
The catalog contains tables, a set of figures with the time histories 
and energy spectra, and some statistical distributions.

Table~\ref{Table_basic} lists the main characteristics of the events. The first three 
columns specify burst order numbers, burst names according to the date of 
appearance, and trigger times T$_0$. The burst duration  T$_{90}$ measures the 
duration of the time interval during which 90\% of the total observed counts 
in the energy windows G2+G3 were detected. The next two columns present values of peak fluxes and 
fluences. GRB fluences are calculated for emission observed above 15~keV.
Peak fluxes are calculated using an average spectrum accumulated in the 
time interval around the peak. The seventh column presents the energy 
interval for which fluence and peak flux have been  calculated. The 
last  column presents information about the burst localization. The 
events localized by BATSE are marked ``B'' with the trigger number 
(Paciesas et al. 1999). The bursts localized by the IPN have the 
references on the Gamma-Ray Burst Coordinates Network (GCN) Circulars with 
the list of spacecraft or experiments which detected the burst 
(U---Ulysses, K---Konus (``Wind''), N---NEAR).

Table~\ref{Table_specpar} contains the spectral data. The first two columns indicate the 
burst names and trigger times. The third  and fourth columns indicate the 
start and stop times of the spectral measurements. 
The fifth column contains the energy interval in which the energy spectra 
were fitted. The sixth column indicates the type of spectral model 
used  for  fitting. The last four columns contain the parameters of spectral
model~\footnote{In case the error of paremeter is not given, the parameter was fixed when fitting}:
A, $\alpha$, $E_0$, $\beta$ (in the notation of section~3).

A set of figures, from Fig.~1 up to  Fig.~164, includes time histories and 
deconvolved background-subtracted photon spectra (panels ``a'' and ``b''), 
respectively, for observed bursts. Time histories are usually given for the 
three energy windows G1, G2, and G3 together with the count rate ratios 
G2/G1 and G3/G2. Energy spectra were accumulated in the indicated time 
intervals. On the time history graphs, these intervals are marked by 
two vertical dotted lines.

\subsection{Afterglows of short bursts}

The early afterglows of short GRBs are observed for 11 of 130 
short bursts. These afterglows appear in time intervals from several 
seconds up to 100 seconds after the trigger. Characteristics of these 
bursts and their afterglows are collected in Tables~\ref{Table_tails_basic} and~\ref{Table_tails_specpar}.

Table~3 contains the data for fluences and peak fluxes. The first three 
columns specify burst names according to the date of appearance, the figure 
number on which time history and spectral data are displayed, and trigger 
times T$_0$. In the next columns, one of two lines presents characteristics of 
the burst. The other line contains characteristics of its afterglow. The 
energy intervals in which parameters of the burst and its afterglow have 
been determined are shown in the fourth column. The following columns 
present  the time intervals of the fluence measurements, 
the  fluences, the time intervals of the peak flux measurements and the peak 
fluxes.

Table~\ref{Table_tails_specpar} summarizes the spectral parameters of the bursts and their afterglows. 
Its first columns present burst names and trigger times T$_0$, 
starting and ending times of the spectral measurements, and energy range 
covered. The next columns indicate the spectral model used  for  fitting, and 
parameters $\alpha$, $\beta$ and $E_0$ (in the notation of section~3).
The multichannel energy spectra were obtained only in the initial stage of the afterglow for 
five of the eleven events in Table~\ref{Table_tails_specpar}.
These spectra are shown in Figures~93, 
99, 103, 117 and 131. In the other cases, the spectra were too weak for meaningful
values to be obtained. In these cases, accumulated count rates in the energy windows 
G1, G2, and G3 were used to estimate spectra using the procedure developed for the 
analysis of fast spectral variability of GRB emission 
(Mazets et~al. 2000). These estimates are also given in Table~\ref{Table_tails_specpar}.

An early afterglow search for all the other short bursts has been carried out
using the background mode data. Summing the time histories for numbers of events
has permitted improved statistics which reveal the existence of an afterglow in 60 per 
cent of the strongest events at time interval up to 250 seconds after the burst 
trigger. This results, suggesting that an early afterglow is a 
typical feature of short bursts will be considered further elsewhere.

Here we discuss only two examples, which show that the distinguishing between
short bursts with an afterglow and the more common long bursts may not always be easy.
The event GRB~000727 begins with a 
typical short hard peak, but 7 seconds later a second short peak appears 
with a power law energy spectrum that is typical of the final stage in the
evolution of long bursts. GRB~000218 may be either a short burst with 
an afterglow or a long burst with an intense initial stage.

\section{CONCLUDING REMARKS}
Some statistical distributions of the main characteristics of the short GRBs
are presented in Figures~165--172.
The electronic version of the catalog is available at \url{http://www.ioffe.ru/LEA/shortGRBs/Catalog/}.
It contains detailed information about the characteristics of
the short GRBs archived as ASCII files.

This work was supported by Russian Aviation and Space Agency Contract, and 
RFBR grant N~01-02-17808.

\clearpage
\begin{deluxetable}{llrrcclc}
\tabletypesize{\footnotesize}
\tablecaption{Short GRBs: basic data\label{Table_basic}}
\tablehead{
\colhead{N} & \colhead{Burst}&\colhead{T$_0$}& \colhead{T$_{90}$} & \colhead{Peak Flux} &%
\colhead{Fluence} & \colhead{Energy interval} & \colhead{Localization} \\
& \colhead{name}& \colhead{s UT}& \colhead{s} & \colhead{erg cm$^{-2}$ s$^{-1}$} &%
\colhead{erg cm$^{-2}$} & \colhead{keV} &%
}
\startdata
1 & 950210 & 8424.148 & 0.14 & 7.2$\times 10^{-6}$ & 7.1$\times 10^{-7}$ & 15--2000 & B (tr. 3410) \\
2 & 950211a & 8697.749 & 0.12 & 6.4$\times 10^{-5}$ & 3.0$\times 10^{-6}$ & 15--5000 & B (tr. 3412) \\
3 & 950414 & 40882.798 & 0.14 & 1.2$\times 10^{-5}$ & 8.2$\times 10^{-7}$ & 15--1000 & \nodata \\
4 & 950419a & 8628.860 & 1.04 & 3.5$\times 10^{-5}$ & 1.8$\times 10^{-5}$ & 15--5000 & \nodata \\
5 & 950520 & 83271.404 & 1.10 & 9.5$\times 10^{-6}$ & 2.0$\times 10^{-6}$ & 15--5000 & \nodata \\
6 & 950610b & 19096.034 & 0.07 & 1.6$\times 10^{-5}$ & 7.1$\times 10^{-7}$ & 15--1000 & \nodata \\
7 & 950726 & 51579.299 & 0.96 & 2.4$\times 10^{-6}$ & 1.7$\times 10^{-6}$ & 15--5000 & B (tr. 3709) \\
8 & 950805b & 13454.144 & 0.61 & 3.4$\times 10^{-5}$ & 1.8$\times 10^{-6}$ & 15--5000 & B (tr. 3736) \\
9 & 951013 & 57097.299 & 0.04 & 1.6$\times 10^{-5}$ & 4.4$\times 10^{-7}$ & 15--1000 & \nodata \\
10 & 951014a & 13108.167 & 1.51 & 7.3$\times 10^{-5}$ & 2.9$\times 10^{-5}$ & 15--5000 & \nodata \\
11 & 960312 & 51579.299 & 0.19 & 2.1$\times 10^{-6}$ & 4.1$\times 10^{-7}$ & 15--1000 & \nodata \\
12 & 960319 & 51992.828 & 0.36 & 1.7$\times 10^{-5}$ & 2.7$\times 10^{-6}$ & 15--1000 & B (tr. 5277) \\
13 & 960420 & 17324.809 & 0.30 & 9.6$\times 10^{-6}$ & 1.5$\times 10^{-6}$ & 15--1000 & \nodata \\
14 & 960519 & 15966.283 & 0.60 & 1.3$\times 10^{-4}$ & 1.2$\times 10^{-5}$ & 15--3000 & \nodata \\
15 & 960602 & 42664.032 & 0.28 & 9.6$\times 10^{-6}$ & 1.6$\times 10^{-6}$ & 15--1000 & \nodata \\
16 & 960610 & 84502.254 & 0.47 & 2.3$\times 10^{-5}$ & 5.0$\times 10^{-6}$ & 15--6000 & \nodata \\
17 & 960614 & 67654.516 & 0.12 & 2.3$\times 10^{-5}$ & 2.7$\times 10^{-6}$ & 15--2000 & \nodata \\
18 & 960803 & 67525.033 & 0.05 & 2.1$\times 10^{-5}$ & 8.9$\times 10^{-7}$ & 15--5000 & B (tr. 5561) \\
19 & 960902 & 58097.128 & 1.57 & 1.0$\times 10^{-5}$ & 5.0$\times 10^{-6}$ & 15--2000 & \nodata \\
20 & 960908 & 25028.442 & 0.36 & 6.7$\times 10^{-5}$ & 1.2$\times 10^{-5}$ & 15--5000 & \nodata \\
21 & 961113 & 80522.580 & 0.25 & 1.1$\times 10^{-5}$ & 9.8$\times 10^{-7}$ & 15--1000 &  \nodata \\
22 & 961123 & 80522.580 & 0.26 & 1.8$\times 10^{-5}$ & 1.6$\times 10^{-6}$ & 15--1000 & \nodata \\
23 & 961212 & 14870.487 & 0.86 & 6.5$\times 10^{-5}$ & 2.7$\times 10^{-5}$ & 15--8000 & B (tr. 5711) \\
24 & 970222 & 86006.565 & 0.85 & 7.8$\times 10^{-5}$ & 8.9$\times 10^{-6}$ & 15--3000 & \nodata \\
25 & 970315b & 30064.853 & 0.13 & 2.0$\times 10^{-5}$ & 1.5$\times 10^{-6}$ & 15--1000 & B (tr. 6123) \\
26 & 970330 & 43988.805 & 0.52 & 7.9$\times 10^{-6}$ & 2.3$\times 10^{-6}$ & 15--1000 & \nodata \\
27 & 970427 & 45723.327 & 0.06 & 1.0$\times 10^{-5}$ & 3.9$\times 10^{-7}$ & 15--1000 & B (tr. 6211) \\
28 & 970428 & 13365.268 & 0.62 & 2.8$\times 10^{-5}$ & 4.9$\times 10^{-6}$ & 15--3000 & \nodata \\
28 & 970506 & 56603.264 & 1.59 & 5.5$\times 10^{-6}$ & 4.5$\times 10^{-6} $ & 15--1000 & \nodata \\
31 & 970521 & 49991.214 & 0.31 & 1.5$\times 10^{-5}$ & 2.0$\times 10^{-6}$ & 15--3000 & \nodata \\
30 & 970608 & 49032.954 & 0.98 & 7.7$\times 10^{-6}$ & 4.0$\times 10^{-6} $ & 15--2000 & \nodata \\
32 & 970625a & 23681.548 & 0.03 & 3.8$\times 10^{-5}$ & 9.4$\times 10^{-7}$ & 15--3000 & \nodata \\
33 & 970626 & 6239.033 & 0.13 & 1.7$\times 10^{-5}$ & 1.8$\times 10^{-6}$ & 15--5000 & \nodata \\
34 & 970704 & 4097.025 & 0.09 & 1.5$\times 10^{-3}$ & 4.2$\times 10^{-5}$ & 15--10000 & B (tr. 6293) \\
35 & 970803 & 66535.704 & 0.15 & 8.4$\times 10^{-6}$ & 9.5$\times 10^{-7}$ & 15--1000 & B (tr. 6325) \\
36 & 970902a & 27561.329 & 0.44 & 1.3$\times 10^{-5}$ & 3.6$\times 10^{-6}$ & 15--5000 & \nodata \\
37 & 970921 & 83828.200 & 0.06 & 1.0$\times 10^{-4}$ & 4.5$\times 10^{-6}$ & 15--8000 & \nodata \\
38 & 971015 & 30459.796 & 0.12 & 1.9$\times 10^{-5}$ & 1.7$\times 10^{-6}$ & 15--2000 & B (tr. 6436) \\
39 & 971031 & 23420.942 & 0.16 & 8.0$\times 10^{-6}$ & 1.3$\times 10^{-6}$ & 15--1000 & \nodata \\
40 & 971118 & 29008.529 & 0.08 & 2.7$\times 10^{-6}$ & 1.7$\times 10^{-7}$ & 15--1000 & B (tr. 6486) \\
41 & 971218b & 52503.029 & 1.06 & 6.7$\times 10^{-6}$ & 5.1$\times 10^{-6}$ & 15--2000 & B (tr. 6535) \\
42 & 971230 & 83750.187 & 0.32 & 1.8$\times 10^{-5}$ & 1.7$\times 10^{-6}$ & 15--5000 & \nodata \\
43 & 980205 & 19785.239 & 1.42 & 3.7$\times 10^{-4}$ & 5.9$\times 10^{-6}$ & 15--5000 & \nodata \\
44 & 980218b & 54768.157 & 0.28 & 4.5$\times 10^{-6}$ & 1.2$\times 10^{-6}$ & 15--2000 & B (tr. 6606) \\
45 & 980228a & 24244.602 & 0.45 & 3.2$\times 10^{-5}$ & 7.6$\times 10^{-6}$ & 15--2000 & B (tr. 6617) \\
46 & 980302b & 29955.993 & 0.30 & 6.2$\times 10^{-6}$ & 1.1$\times 10^{-6}$ & 15--1000 & \nodata \\
47 & 980310a & 50261.054 & 0.80 & 5.3$\times 10^{-6}$ & 2.0$\times 10^{-6}$ & 15--2000 & B (tr. 6635) \\
48 & 980330a & 96.711 & 0.10 & 5.2$\times 10^{-5}$ & 3.0$\times 10^{-6}$ & 15--5000 & B (tr. 6668) \\
49 & 980331 & 61078.449 & 0.17 & 6.7$\times 10^{-6}$ & 1.1$\times 10^{-6}$ & 15--2000 & B (tr. 6671) \\
50 & 980429 & 20492.079 & 0.18 & 1.7$\times 10^{-5}$ & 1.8$\times 10^{-6}$ & 15--2000 & \nodata \\
51 & 980430 & 59702.214 & 0.30 & 3.2$\times 10^{-5}$ & 6.5$\times 10^{-6}$ & 15--4000 & \nodata \\
52 & 980605 & 51131.976 & 0.13 & 9.4$\times 10^{-5}$ & 2.5$\times 10^{-6}$ & 15--5000 & \nodata \\
53 & 980610a & 71546.850 & 0.86 & 6.0$\times 10^{-6}$ & 3.3$\times 10^{-6}$ & 15--4000 & \nodata \\
54 & 980610b & 86195.164 & 0.31 & 1.5$\times 10^{-5}$ & 3.0$\times 10^{-6}$ & 15--4000 & \nodata \\
55 & 980619 & 47530.372 & 0.11 & 7.0$\times 10^{-5}$ & 3.5$\times 10^{-6}$ & 15--4000 & \nodata \\
56 & 980706a & 57586.277 & 0.32 & 1.7$\times 10^{-4}$ & 4.0$\times 10^{-5}$ & 15--6000 & B (tr. 6904) \\
57 & 980904 & 31349.014 & 0.05 & 3.5$\times 10^{-5}$ & 1.3$\times 10^{-6}$ & 15--2000 & B (tr. 7063) \\
58 & 980908 & 82263.835 & 0.45 & 5.5$\times 10^{-6}$ & 1.1$\times 10^{-6}$ & 15--1500 & \nodata \\
59 & 980925a & 17571.284 & 0.45 & 4.4$\times 10^{-6}$ & 1.7$\times 10^{-6}$ & 15--1500 & \nodata \\
60 & 981005 & 64826.466 & 0.84 & 5.7$\times 10^{-6}$ & 2.8$\times 10^{-6}$ & 15--2000 & B (tr. 7142) \\
61 & 981102 & 28554.533 & 0.58 & 4.2$\times 10^{-5}$ & 1.5$\times 10^{-5}$ & 15--4000 & \nodata \\
62 & 981107 & 781.395 & 0.37 & 6.7$\times 10^{-4}$ & 9.9$\times 10^{-5}$ & 15--8000 & \nodata \\
63 & 981218 & 62134.933 & 0.30 & 2.8$\times 10^{-6}$ & 7.7$\times 10^{-7}$ & 15--2000 & \nodata \\
64 & 981221 & 9057.150 & 0.13 & 9.4$\times 10^{-6}$ & 8.5$\times 10^{-7}$ & 15--1000 & B (tr. 7273) \\
65 & 981226 & 38822.991 & 0.64 & 6.1$\times 10^{-6}$ & 2.7$\times 10^{-6}$ & 15--1000 & B (tr. 7281) \\
66 & 990105 & 31789.507 & 0.52 & 1.2$\times 10^{-5}$ & 1.9$\times 10^{-6}$ & 15--1000 & B (tr. 7305) \\
67 & 990126 & 51844.333 & 0.23 & 3.6$\times 10^{-5}$ & 3.0$\times 10^{-6}$ & 15--2000 & B (tr. 7353) \\
68 & 990206c & 57061.328 & 0.17 & 7.6$\times 10^{-6}$ & 7.7$\times 10^{-7}$ & 15--1000 & B (tr. 7375) \\
69 & 990207 & 69675.009 & 0.39 & 1.8$\times 10^{-5}$ & 1.9$\times 10^{-6}$ & 15--1500 & \nodata \\
70 & 990208 & 15166.066 & 0.53 & 3.1$\times 10^{-6}$ & 8.5$\times 10^{-7}$ & 15--1000 & B (tr. 7378) \\
71 & 990313 & 33712.652 & 0.27 & 1.8$\times 10^{-5}$ & 1.5$\times 10^{-6}$ & 15--1000 & \nodata \\
72 & 990327 & 22911.102 & 0.08 & 1.6$\times 10^{-4}$ & 1.4$\times 10^{-5}$ & 15--5000 & \nodata \\
73 & 990405b & 30059.858 & 0.49 & 4.8$\times 10^{-6}$ & 8.3$\times 10^{-7}$ & 15--2000 & \nodata \\
74 & 990415 & 2297.309 & 0.17 & 7.1$\times 10^{-6}$ & 1.2$\times 10^{-6}$ & 15--1000 & \nodata \\
75 & 990504 & 67586.484 & 0.08 & 1.5$\times 10^{-5}$ & 1.2$\times 10^{-6}$ & 15--2000 & \nodata \\
76 & 990516 & 86065.136 & 0.31 & 2.1$\times 10^{-4}$ & 5.8$\times 10^{-5}$ & 15--5000 & B (tr. 7569) \\
77 & 990619 & 46930.367 & 1.53 & 3.8$\times 10^{-6}$ & 1.8$\times 10^{-6}$ & 15--1500 & B (tr. 7610) \\
78 & 990712a & 27915.510 & 0.64 & 5.3$\times 10^{-5}$ & 2.5$\times 10^{-5}$ & 15--3000 & B (tr. 7647) \\
79 & 990719 & 61135.420 & 0.06 & 1.0$\times 10^{-5}$ & 5.4$\times 10^{-7}$ & 15--1000 & \nodata \\
80 & 990720 & 75941.940 & 0.87 & 3.5$\times 10^{-6}$ & 1.8$\times 10^{-6}$ & 15--1000 & B (tr. 7663) \\
81 & 990806b & 60168.676 & 0.14 & 9.3$\times 10^{-6}$ & 1.1$\times 10^{-6}$ & 15--1000 & \nodata \\
82 & 990828 & 70020.016 & 1.12 & 5.9$\times 10^{-6}$ & 2.1$\times 10^{-6}$ & 15--500 & \nodata \\
83 & 990831 & 41835.091 & 0.08 & 5.9$\times 10^{-6}$ & 3.8$\times 10^{-7}$ & 15--1000 & \nodata \\
84 & 991001 & 4950.128 & 0.82 & 2.9$\times 10^{-6}$ & 2.2$\times 10^{-6}$ & 15--1000 & B (tr. 7781) \\
85 & 991002 & 82143.664 & 0.25 & 3.7$\times 10^{-6}$ & 7.4$\times 10^{-7}$ & 15--1000 & B (tr. 7784) \\
86 & 991226b & 83339.767 & 0.08 & 7.7$\times 10^{-6}$ & 3.9$\times 10^{-7}$ & 15--1000 & \nodata \\
87 & 000108 & 60487.439 & 0.72 & 3.1$\times 10^{-6}$ & 2.2$\times 10^{-6}$ & 15--1000 & B (tr. 7939) \\
88 & 000212 & 61592.339 & 0.17 & 4.3$\times 10^{-6}$ & 5.2$\times 10^{-7}$ & 15--1000 & \nodata \\
89 & 000218 & 58744.596 & 0.90 & 8.5$\times 10^{-5}$ & 4.7$\times 10^{-5}$ & 15--6000 & \nodata \\
90 & 000326 & 19134.798 & 0.96 & 5.2$\times 10^{-6}$ & 3.5$\times 10^{-6}$ & 15--1000 & IPN (U-K-N; GCN 618), B (tr. 8053) \\
91 & 000412 & 42174.189 & 0.52 & 2.0$\times 10^{-6}$ & 5.7$\times 10^{-7}$ & 15--1000 & B (tr. 8073) \\
92 & 000420a & 42271.144 & 0.20 & 1.6$\times 10^{-5}$ & 2.1$\times 10^{-6}$ & 15--3000 & \nodata \\
93 & 000513 & 40894.793 & 0.35 & 4.1$\times 10^{-6}$ & 8.1$\times 10^{-7}$ & 15--1000 & B (tr. 8104) \\
94 & 000526 & 84494.896 & 0.27 & 8.9$\times 10^{-5}$ & 1.5$\times 10^{-5}$ & 15--5000 & \nodata \\
95 & 000607 & 8689.115 & 0.09 & 1.2$\times 10^{-4}$ & 5.4$\times 10^{-6}$ & 15--3000 & IPN (U-K-N; GCN 693) \\
96 & 000608 & 70497.255 & 0.18 & 1.7$\times 10^{-5}$ & 2.7$\times 10^{-6}$ & 15--1000 & \nodata \\
97 & 000623 & 3887.359 & 0.17 & 5.9$\times 10^{-6}$ & 8.1$\times 10^{-7}$ & 15--3000 & \nodata \\
98 & 000701b & 25961.013 & 0.94 & 2.6$\times 10^{-5}$ & 1.3$\times 10^{-5}$ & 15--3000 & \nodata \\
99 & 000707a & 17372.072 & 0.21 & 5.1$\times 10^{-6}$ & 8.4$\times 10^{-7}$ & 15--1000 & \nodata \\
100 & 000727 & 70955.931 & 0.80 & 5.2$\times 10^{-5}$ & 1.9$\times 10^{-5}$ & 15--1000 & IPN (U-K-N; GCN 754) \\
101 & 000818 & 72547.04 & 1.35 & 5.6$\times 10^{-6}$ & 2.3$\times 10^{-6}$ & 15--1000 & \nodata \\
102 & 000928 & 6285.374 & 0.22 & 2.4$\times 10^{-5}$ & 4.0$\times 10^{-6}$ & 15--3000 & \nodata \\
103 & 001022 & 20905.666 & 0.12 & 5.0$\times 10^{-5}$ & 3.5$\times 10^{-6}$ & 15--3000 & \nodata \\
104 & 001025c & 71369.963 & 0.48 & 2.2$\times 10^{-5}$ & 5.9$\times 10^{-6}$ & 15--5000 & IPN (U-N-K; GCN 865) \\
105 & 001204 & 28869.372 & 0.27 & 9.0$\times 10^{-6}$ & 1.7$\times 10^{-6}$ & 15--5000 & IPN (U-N-S-K; GCN 895, 897) \\
106 & 001207a & 8815.958 & 0.17 & 7.7$\times 10^{-6}$ & 7.4$\times 10^{-7}$ & 15--1000 & \nodata \\
107 & 001207b & 34185.588 & 0.83 & 2.5$\times 10^{-6}$ & 2.2$\times 10^{-6}$ & 15--1000 & \nodata \\
108 & 010119 & 37179.556 & 0.18 & 3.6$\times 10^{-5}$ & 2.4$\times 10^{-6}$ & 15--5000 & IPN (U-N-K; GCN 916) \\
109 & 010308 & 56338.468 & 0.85 & 9.6$\times 10^{-6}$ & 5.3$\times 10^{-6}$ & 15--2000 & \nodata \\
110 & 010420a & 30786.674 & 0.01 & 1.2$\times 10^{-5}$ & 2.3$\times 10^{-7}$ & 15--1000 & \nodata \\
111 & 010427 & 67452.969 & 0.69 & 2.6$\times 10^{-5}$ & 5.4$\times 10^{-6}$ & 15--3000 & \nodata \\
112 & 010616 & 23724.028 & 0.10 & 1.8$\times 10^{-5}$ & 1.5$\times 10^{-6}$ & 15--1000 & \nodata \\
113 & 010624 & 48929.130 & 0.64 & 3.8$\times 10^{-6}$ & 1.7$\times 10^{-6}$ & 15--1000 & \nodata \\
114 & 010628a & 4206.816 & 0.96 & 4.1$\times 10^{-5}$ & 1.3$\times 10^{-5}$ & 15--5000 & \nodata \\
115 & 011024 & 74609.296 & 0.61 & 4.1$\times 10^{-6}$ & 2.9$\times 10^{-6}$ & 15--1000 & \nodata \\
116 & 011101 & 34754.534 & 0.46 & 4.8$\times 10^{-6}$ & 2.2$\times 10^{-5}$ & 15--3000 & \nodata \\
117 & 020116 & 78766.919 & 0.08 & \nodata & \nodata & \nodata & \nodata \\
118 & 020117 & 67773.429 & 0.47 & 2.0$\times 10^{-4}$ & 2.7$\times 10^{-4}$ & 15--1000 & \nodata \\
119 & 020218a & 31263.113 & 0.09 & \nodata & \nodata & \nodata & \nodata \\
120 & 020306 & 68280.713 & 0.15 & 3.8$\times 10^{-5}$ & 4.7$\times 10^{-6}$ & 15--6000 & \nodata \\
121 & 020326 & 39182.941 & 0.12 & 8.8$\times 10^{-6}$ & 9.3$\times 10^{-7}$ & 15--1000 & \nodata \\
122 & 020426 & 86171.048 & 0.19 & \nodata  & \nodata  & \nodata & \nodata \\
123 & 020504 & 55835.141 & 1.05 & \nodata & 2.8$\times 10^{-5}$ & 15--5000 & \nodata \\
124 & 020509 & 74.563 & 1.08 & \nodata & \nodata & \nodata & \nodata \\
125 & 020525a & 16014.630 & 0.18 & 1.5$\times 10^{-5}$  & 2.0$\times 10^{-6}$ & 15--2000 & \nodata \\
126 & 020602b & 63030.315 & 0.46 & 8.0$\times 10^{-6}$ & 1.1$\times 10^{-6}$ & 15--1000 & \nodata \\
127 & 020715a & 54866.135 & 0.43 & 9.0$\times 10^{-6}$ & 1.4$\times 10^{-6}$ & 15--2000 & \nodata \\
128 & 020731a & 1635.905 & 0.12 & 2.4$\times 10^{-5}$ & 3.1$\times 10^{-6}$ & 15--2000 & \nodata \\
129 & 020731b & 50231.739 & 0.29 & 3.4$\times 10^{-6}$ & 9.5$\times 10^{-7}$ & 15--1000 & \nodata \\
130 & 020828 & 20737.981 & 0.70 & 1.1$\times 10^{-5}$ & 4.0$\times 10^{-6}$ & 15--3000 & \nodata \\
\enddata
\end{deluxetable}
\begin{deluxetable}{lrrrlllrll}
\tabletypesize{\scriptsize}
\tablecaption{Short GRBs: spectral data\label{Table_specpar}}
\tablehead{
\colhead{Burst}&\colhead{T$_0$}& \multicolumn{2}{c}{Time interval} &%
\colhead{Energy} & \colhead{Model\tablenotemark{1}} & \colhead{A} & \colhead{$\alpha$} & \colhead{$E_0$} & \colhead{$\beta$}\\
\colhead{name}& & \colhead{Start} & \colhead{Stop} &%
\colhead{interval} & & \colhead{photons} & & &\\
& \colhead{s UT} & \colhead{s} & \colhead{s} & \colhead{keV} & & cm$^{-2}$ s$^{-1}$ keV$^{-1}$ & & \colhead{keV} & 
}
\startdata
950210 & 8424.148 & 0 & 0.128 & 15--2000 & GRB & $(3.8 \pm 0.5) \times 10^{-1}$ & -0.20 & $(7.4 \pm 0.7) \times 10^{1}$ & -2.95 $\pm$ 0.58 \\
950211a & 8697.749 & 0 & 0.128 & 15--5000 & GRB & $(3.4 \pm 0.5) \times 10^{-1}$ & -0.31 $\pm$ 0.14 & $(1.7 \pm 0.3) \times 10^{2}$ & -2.69 $\pm$ 0.38 \\
950419a & 8628.860 & 0 & 0.256 & 15--5000 & GRB & $(1.7 \pm 0.1) \times 10^{-1}$ & 0.02 $\pm$ 0.17 & $(2.4 \pm 0.4) \times 10^{2}$ & -2.69 $\pm$ 0.38 \\
 & & 0.256 & 8.448 & 15--5000 & GRB & $(1.8 \pm 0.3) \times 10^{-2}$ & -0.29 $\pm$ 0.24 & $(1.9 \pm 0.4) \times 10^{2}$ & -2.70 \\
950520 & 83271.404 & 0 & 0.064 & 15--5000 & COMP & $(7.3 \pm 7.7) \times 10^{-2}$ & -0.70 $\pm$ 0.25 & $(4.5 \pm 2.1) \times 10^{2}$ & \nodata \\
950726 & 51579.299 & 0 & 0.256 & 15--5000 & COMP & $(4.0 \pm 3.9) \times 10^{-2}$ & -0.99 $\pm$ 0.27 & $(2.4 \pm 1.2) \times 10^{2}$ & \nodata \\
950805b & 13454.144 & 0 & 0.064 & 15--5000 & COMP & $(9.6 \pm 5.3) \times 10^{-2}$ & -0.93 $\pm$ 0.13 & $(1.1 \pm 0.5) \times 10^{3}$ & \nodata \\
951014a & 13108.167 & 0 & 0.256 & 15--5000 & GRB & $(3.6 \pm 0.4) \times 10^{-1}$ & -0.17 $\pm$ 0.13 & $(2.1 \pm 0.3) \times 10^{2}$ & -2.12 $\pm$ 0.12 \\
 & & 0.256 & 0.512 & 15--5000 & GRB & $(4.8 \pm 0.4) \times 10^{-1}$ & -0.55 $\pm$ 0.10 & $(2.4 \pm 0.4) \times 10^{2}$ & -2.25 $\pm$ 0.14 \\
 & & 0.512 & 7.936 & 15--3000 & GRB & $(1.8 \pm 0.2) \times 10^{-2}$ & -1.36 $\pm$ 0.12 & $(4.5 \pm 1.8) \times 10^{2}$ & -2.15 $\pm$ 0.10 \\
960319 & 51992.828 & 0 & 0.256 & 15--3000 & COMP & $(2.1 \pm 1.7) \times 10^{-2}$ & -0.93 $\pm$ 0.17 & $(1.6 \pm 1.3) \times 10^{3}$ & \nodata \\
960519 & 14766.283 & 0 & 8.448 & 15--3000 & PL & $(3.5 \pm 1.5) \times 10^{-3}$ & -1.24 $\pm$ 0.07 & \nodata & \nodata \\
960610 & 84502.254 & 0 & 0.128 & 15--6000 & COMP & $(4.1 \pm 0.5) \times 10^{-2}$ & -0.50 & $(9.4 \pm 1.6) \times 10^{2}$ & \nodata \\
960614 & 67654.516 & 0 & 0.064 & 15--2000 & COMP & $(8.0 \pm 4.6) \times 10^{-2}$ & -0.79 $\pm$ 0.14 & $1.0 \times 10^{3}$ & \nodata \\
960803 & 67525.033 & 0 & 0.064 & 15--5000 & GRB & $(5.2 \pm 7.4) \times 10^{-1}$ & -0.28 $\pm$ 0.82 & $(5.8 \pm 5.2) \times 10^{1}$ & -2.05 $\pm$ 0.23 \\
960902 & 58097.128 & 0 & 0.128 & 15--2000 & GRB & $(2.1 \pm 1.1) \times 10^{-1}$ & -0.71 $\pm$ 0.38 & $(1.4 \pm 1.0) \times 10^{2}$ & -1.95 $\pm$ 0.18 \\
 & & 0.256 & 7.936 & 15--500 & COMP & $(4.3 \pm 3.6) \times 10^{-2}$ & -0.67 $\pm$ 0.26 & $(6.7 \pm 1.5) \times 10^{1}$ & \nodata \\
960908 & 25028.442 & 0 & 0.128 & 15--3000 & PL & $(1.3 \pm 0.2) \times 10^{-1}$ & -1.32 $\pm$ 0.03 & \nodata & \nodata \\
 & & 0.128 & 0.256 & 15--3000 & PL & $(1.7 \pm 0.2) \times 10^{-1}$ & -1.62 $\pm$ 0.03 & \nodata & \nodata \\
961212 & 14870.487 & 0 & 0.512 & 15--8000 & GRB & $(1.7 \pm 0.1) \times 10^{-1}$ & -0.89 $\pm$ 0.07 & $(4.7 \pm 0.9) \times 10^{2}$ & -2.05 \\
 & & 0.512 & 8.704 & 15--5000 & COMP & $(7.3 \pm 1.5) \times 10^{-3}$ & -1.58 $\pm$ 0.04 & $1.0 \times 10^{3}$ & \nodata \\
970222 & 86006.565 & 0 & 0.128 & 15--3000 & GRB & $(1.7 \pm 0.4) \times 10^{-1}$ & -0.51 $\pm$ 0.21 & $(2.4 \pm 1.0) \times 10^{2}$ & -1.85 $\pm$ 0.21 \\
 & & 0.256 & 8.448 & 15--3000 & GRB & $(3.5 \pm 0.3) \times 10^{-3}$ & -0.99 $\pm$ 0.09 & $3.6 \times 10^{2}$ & -2.30 \\
970428 & 13365.268 & 0 & 0.256 & 15--5000 & COMP & $(6.1 \pm 2.6) \times 10^{-2}$ & -0.82 $\pm$ 0.10 & $(5.8 \pm 1.4) \times 10^{2}$ & \nodata \\
 & & 0.256 & 8.448 & 15--5000 & PL & $(2.6 \pm 1.1) \times 10^{-3}$ & -1.29 $\pm$ 0.08 & \nodata & \nodata \\
970506 & 56603.264 & 0 & 0.256 & 15--1000 & GRB & 1.2 $\pm$ 1.5 & 0.53 $\pm$ 0.87 & $(3.8 \pm 1.8) \times 10^{1}$ & -2.64 $\pm$ 0.25 \\
 & & 0.256 & 8.448 & 15--1000 & PL & $(5.4 \pm 1.8) \times 10^{-3}$ & -1.88 $\pm$ 0.08 & \nodata & \nodata \\
970521 & 49991.214 & 0 & 0.256 & 15--3000 & COMP & $(5.1 \pm 5.5) \times 10^{-2}$ & -0.89 $\pm$ 0.25 & $(4.8 \pm 2.3) \times 10^{2}$ & \nodata \\
970608 & 49032.954 & 0 & 0.256 & 15--2000 & COMP & $(2.3 \pm 2.0) \times 10^{-2}$ & -0.88 $\pm$ 0.19 & $(1.5 \pm 1.2) \times 10^{3}$ & \nodata \\
970625a & 23681.548 & 0 & 0.064 & 15--3000 & COMP & $(6.0 \pm 3.2) \times 10^{-2}$ & -1.32 $\pm$ 0.10 & $1.5 \times 10^{3}$ & \nodata \\
970626 & 6239.033 & 0 & 0.128 & 15--5000 & COMP & $(5.3 \pm 4.5) \times 10^{-2}$ & -1.14 $\pm$ 0.19 & $(1.4 \pm 1.2) \times 10^{3}$ & \nodata \\
970704 & 4097.025 & 0 & 0.064 & 15--10000 & COMP & $(5.3 \pm 1.4) \times 10^{-1}$ & -0.76 $\pm$ 0.05 & $(2.5 \pm 0.4) \times 10^{3}$ & \nodata \\
 & & 0.064 & 0.128 & 15--5000 & GRB & $(8.2 \pm 1.5) \times 10^{-1}$ & -0.18 $\pm$ 0.20 & $(1.5 \pm 0.3) \times 10^{2}$ & -2.72 $\pm$ 0.34 \\
 & & 0.128 & 0.192 & 15--1000 & COMP & $(2.9 \pm 3.4) \times 10^{-1}$ & -0.82 $\pm$ 0.39 & $(5.7 \pm 2.0) \times 10^{1}$ & \nodata \\
970902a & 27561.329 & 0 & 0.256 & 15--5000 & COMP & $(5.3 \pm 3.4) \times 10^{-2}$ & -0.87 $\pm$ 0.14 & $(8.1 \pm 3.0) \times 10^{2}$ & \nodata \\
970921 & 83828.200 & 0 & 0.064 & 15--8000 & GRB & $(3.8 \pm 0.5) \times 10^{-1}$ & 0.14 $\pm$ 0.27 & $(2.1 \pm 0.5) \times 10^{2}$ & -2.30 \\
971015 & 30459.796 & 0 & 0.064 & 15--2000 & COMP & $(7.4 \pm 4.5) \times 10^{-2}$ & -1.08 $\pm$ 0.15 & $(1.7 \pm 1.6) \times 10^{3}$ & \nodata \\
971218b & 52503.029 & 0 & 0.256 & 15--2000 & PL & $(3.8 \pm 1.5) \times 10^{-2}$ & -1.45 $\pm$ 0.09 & \nodata & \nodata \\
 & & 0.256 & 8.448 & 15--1000 & PL & $(2.4 \pm 1.2) \times 10^{-3}$ & -1.42 $\pm$ 0.11 & \nodata & \nodata \\
971230 & 83750.187 & 0 & 0.128 & 15--2000 & PL & $(2.1 \pm 1.1) \times 10^{-2}$ & -1.18 $\pm$ 0.09 & \nodata & \nodata \\
980205\tablenotemark{2} & 19785.239 & 0 & 0.256 & 15--5000 & COMP+ & $(3.5 \pm 2.0) \times 10^{-2}$ & -0.88 $\pm$ 0.09 & $2.0 \times 10^3$ & \nodata \\
 & & & &				& PL & $(4.6 \pm 0.3) \times 10^{-2}$ & -2.3 & \nodata & \nodata \\
 &  & 0.256 & 8.448 & 15--1000 & PL & $(1.3 \pm 1.1) \times 10^{-3}$ & -2.21 $\pm$ 0.23 & \nodata & \nodata \\
980228a & 24244.602 & 0 & 0.256 & 15--5000 & GRB & $(6.9 \pm 0.8) \times 10^{-1}$ & -0.53 $\pm$ 0.13 & $(1.5 \pm 0.2) \times 10^{2}$ & -3.46 $\pm$ 0.46 \\
 & & 0.256 & 8.448 & 15--2000 & COMP & $(7.0 \pm 6.3) \times 10^{-3}$ & -1.71 $\pm$ 0.24 & $(5.9 \pm 5.5) \times 10^{2}$ & \nodata \\
980310a & 50261.054 & 0 & 0.256 & 15--2000 & COMP & $(4.0 \pm 1.3) \times 10^{-2}$ & -1.74 $\pm$ 0.08 & $5.0 \times 10^{2}$ & \nodata \\
 & & 0.256 & 8.448 & 15--2000 & COMP & $(4.2 \pm 1.7) \times 10^{-3}$ & -1.80 $\pm$ 0.10 & $5.0 \times 10^{2}$ & \nodata \\
980330a & 96.711 & 0 & 0.064 & 15--5000 & GRB & $(4.4 \pm 0.7) \times 10^{-1}$ & -0.53 $\pm$ 0.21 & $(2.4 \pm 0.7) \times 10^{2}$ & -2.60 $\pm$ 0.45 \\
980331 & 61078.449 & 0 & 0.128 & 15--2000 & COMP & $(3.4 \pm 1.7) \times 10^{-2}$ & -1.18 $\pm$ 0.10 & $1.0 \times 10^{3}$ & \nodata \\
980429 & 20492.079 & 0 & 0.192 & 15--2000 & COMP & $(1.8 \pm 0.1) \times 10^{-1}$ & -0.20 & $(1.7 \pm 0.1) \times 10^{2}$ & \nodata \\
980430 & 59702.214 & 0 & 0.256 & 15--4000 & COMP & $(5.4 \pm 2.2) \times 10^{-2}$ & -0.67 $\pm$ 0.07 & $1.0 \times 10^{3}$ & \nodata \\
980605 & 51131.976 & 0 & 0.064 & 15--5000 & COMP & $(5.0 \pm 5.6) \times 10^{-2}$ & -0.74 $\pm$ 0.25 & $(1.0 \pm 0.8) \times 10^{3}$ & \nodata \\
980610a & 71546.850 & 0 & 0.256 & 15--4000 & COMP & $(3.4 \pm 5.2) \times 10^{-2}$ & -0.18 $\pm$ 0.34 & $(2.9 \pm 1.0) \times 10^{2}$ & \nodata \\
980610b & 86195.164 & 0 & 0.192 & 15--4000 & COMP & $(4.4 \pm 4.5) \times 10^{-2}$ & -1.04 $\pm$ 0.23 & $(1.2 \pm 0.9) \times 10^{3}$ & \nodata \\
980619 & 47530.372 & 0 & 0.064 & 15--4000 & COMP & $(4.1 \pm 0.7) \times 10^{-2}$ & 0.00 & $(5.9 \pm 0.9) \times 10^{2}$ & \nodata \\
980706a & 57586.277 & 0 & 0.256 & 15--6000 & COMP & $(2.6 \pm 0.4) \times 10^{-1}$ & -0.75 $\pm$ 0.03 & $(1.3 \pm 0.1) \times 10^{3}$ & \nodata \\
 & & 0.256 & 8.448 & 15--3000 & PL & $(4.2 \pm 1.2) \times 10^{-3}$ & -1.26 $\pm$ 0.05 & \nodata & \nodata \\
980904 & 31349.014 & 0 & 0.064 & 15--2000 & PL & $(2.6 \pm 0.3) \times 10^{-2}$ & -1.50 & \nodata & \nodata \\
980908 & 82263.835 & 0 & 0.256 & 15--1500 & COMP & $(4.5 \pm 7.0) \times 10^{-2}$ & -0.29 $\pm$ 0.37 & $(1.9 \pm 0.7) \times 10^{2}$ & \nodata \\
980925a & 17571.284 & 0 & 0.256 & 15--1500 & PL & $(3.0 \pm 1.4) \times 10^{-2}$ & -1.85 $\pm$ 0.10 & \nodata & \nodata \\
981005 & 64826.466 & 0 & 0.256 & 15--2000 & COMP & $(1.6 \pm 1.1) \times 10^{-2}$ & -0.95 $\pm$ 0.13 & $8.0 \times 10^{2}$ & \nodata \\
981102 & 28554.533 & 0 & 0.256 & 15--4000 & COMP & $(5.0 \pm 2.8) \times 10^{-2}$ & -0.74 $\pm$ 0.12 & $(1.4 \pm 0.5) \times 10^{3}$ & \nodata \\
981107 & 781.395 & 0 & 0.128 & 15--8000 & COMP & $(8.4 \pm 7.7) \times 10^{-2}$ & -0.08 $\pm$ 0.16 & $(8.0 \pm 1.3) \times 10^{2}$ & \nodata \\
 & & 0.128 & 0.256 & 15--8000 & COMP & $(1.5 \pm 0.6) \times 10^{-1}$ & -0.29 $\pm$ 0.07 & $(1.6 \pm 0.2) \times 10^{3}$ & \nodata \\
 & & 0.256 & 0.512 & 15--8000 & GRB & $(1.9 \pm 0.1) \times 10^{-1}$ & -0.64 $\pm$ 0.04 & $(1.1 \pm 0.1) \times 10^{3}$ & -3.09 $\pm$ 0.60 \\
981218 & 62134.933 & 0 & 0.256 & 15--2000 & COMP & $(2.4 \pm 3.0) \times 10^{-1}$ & -0.09 $\pm$ 0.38 & $(6.1 \pm 1.7) \times 10^{1}$ & \nodata \\
981221 & 9057.150 & 0 & 0.064 & 15--1000 & PL & $(3.2 \pm 2.0) \times 10^{-2}$ & -1.20 $\pm$ 0.11 & \nodata & \nodata \\
981226 & 38822.991 & 0 & 0.256 & 15--1000 & COMP & $(1.1 \pm 0.9) \times 10^{-1}$ & -1.21 $\pm$ 0.21 & $(3.5 \pm 1.5) \times 10^{2}$ & \nodata \\
 & & 0.256 & 8.448 & 15--500 & PL & $(2.5 \pm 0.3) \times 10^{-3}$ & -2.30 $\pm$ 0.50 & \nodata & \nodata \\
990126 & 51844.333 & 0 & 0.064 & 15--2000 & PL & $(3.7 \pm 2.9) \times 10^{-2}$ & -1.05 $\pm$ 0.13 & \nodata & \nodata \\
990207 & 69675.009 & 0 & 0.128 & 15--1500 & COMP & $(6.7 \pm 3.4) \times 10^{-2}$ & -0.66 $\pm$ 0.09 & $5.0 \times 10^{2}$ & \nodata \\
990313 & 33712.652 & 0 & 0.128 & 15--2000 & GRB & $(1.8 \pm 0.5) \times 10^{-1}$ & -1.08 $\pm$ 0.18 & $(2.0 \pm 0.9) \times 10^{2}$ & -2.43 $\pm$ 0.41 \\
990327 & 22911.102 & 0 & 0.064 & 15--5000 & COMP & $(1.7 \pm 0.9) \times 10^{-1}$ & -0.86 $\pm$ 0.11 & $(2.1 \pm 0.8) \times 10^{3}$ & \nodata \\
 & & 0.256 & 16.640 & 15--5000 & COMP & $(6.9 \pm 3.2) \times 10^{-3}$ & -1.39 $\pm$ 0.11 & $(1.2 \pm 0.7) \times 10^{3}$ & \nodata \\
990405b & 30059.858 & 0 & 0.256 & 15--2000 & COMP & $(3.8 \pm 2.0) \times 10^{-2}$ & -1.64 $\pm$ 0.16 & $(4.5 \pm 3.2) \times 10^{2}$ & \nodata \\
990415 & 2297.309 & 0 & 0.128 & 15--1000 & COMP & $(5.6 \pm 9.9) \times 10^{-2}$ & 0.21 $\pm$ 0.40 & $(2.0 \pm 0.6) \times 10^{2}$ & \nodata \\
990504 & 67586.484 & 0 & 0.064 & 15--2000 & COMP & $(8.7 \pm 6.5) \times 10^{-2}$ & -0.94 $\pm$ 0.18 & $(6.6 \pm 3.8) \times 10^{2}$ & \nodata \\
990516 & 86065.136 & 0 & 0.064 & 15--5000 & COMP & $(6.7 \pm 4.5) \times 10^{-2}$ & -1.26 $\pm$ 0.16 & $(1.9 \pm 2.4) \times 10^{3}$ & \nodata \\
 & & 0.128 & 0.256 & 15--6000 & COMP & $(1.7 \pm 0.6) \times 10^{-1}$ & -0.60 $\pm$ 0.07 & $(1.1 \pm 0.2) \times 10^{3}$ & \nodata \\
 & & 0.256 & 8.448 & 15--5000 & PL & $(3.8 \pm 1.5) \times 10^{-3}$ & -1.60 $\pm$ 0.08 & \nodata & \nodata \\
 & & 8.448 & 33.024 & 15--5000 & PL & $(2.8 \pm 0.8) \times 10^{-3}$ & -1.72 $\pm$ 0.06 & \nodata & \nodata \\
990619 & 46930.367 & 0 & 0.256 & 15--1500 & COMP & $(6.0 \pm 6.4) \times 10^{-2}$ & -0.59 $\pm$ 0.27 & $(1.9 \pm 0.6) \times 10^{2}$ & \nodata \\
990712a & 27915.510 & 0 & 0.256 & 15--5000 & COMP & $(6.3 \pm 4.3) \times 10^{-2}$ & -0.20 $\pm$ 0.13 & $(6.1 \pm 1.0) \times 10^{2}$ & \nodata \\
 & & 0.256 & 8.448 & 15--3000 & COMP & $(5.4 \pm 2.2) \times 10^{-3}$ & -0.97 $\pm$ 0.07 & $1.5 \times 10^{3}$ & \nodata \\
 & & 24.832 & 41.216 & 15--1000 & PL & $(3.1 \pm 1.6) \times 10^{-3}$ & -1.98 $\pm$ 0.11 & \nodata & \nodata \\
990719 & 61135.420 & 0 & 0.064 & 15--1000 & COMP & $(1.1 \pm 1.5) \times 10^{-1}$ & -0.76 $\pm$ 0.34 & $(2.2 \pm 1.0) \times 10^{2}$ & \nodata \\
990720 & 75941.940 & 0 & 0.256 & 15--1000 & COMP & $(4.0 \pm 3.1) \times 10^{-2}$ & -1.01 $\pm$ 0.20 & $(4.4 \pm 2.2) \times 10^{2}$ & \nodata \\
990806b & 60168.676 & 0 & 0.128 & 15--1000 & COMP & $(2.4 \pm 4.5) \times 10^{-2}$ & -0.28 $\pm$ 0.38 & $(5.2 \pm 2.6) \times 10^{2}$ & \nodata \\
990828 & 70020.016 & 0 & 0.256 & 15--500 & COMP & $(2.5 \pm 3.1) \times 10^{-1}$ & -0.49 $\pm$ 0.40 & $(5.2 \pm 1.7) \times 10^{1}$ & \nodata \\
 & & 0.256 & 8.448 & 15--500 & PL & $(2.6 \pm 1.1) \times 10^{-3}$ & -1.72 $\pm$ 0.10 & \nodata & \nodata \\
991001 & 4950.128 & 0 & 0.256 & 15--1000 & COMP & $(4.5 \pm 3.3) \times 10^{-2}$ & -1.59 $\pm$ 0.21 & $(4.7 \pm 3.8) \times 10^{2}$ & \nodata \\
 & & 0.256 & 8.448 & 15--1000 & PL & $(3.1 \pm 1.2) \times 10^{-3}$ & -2.03 $\pm$ 0.09 & \nodata & \nodata \\
000108 & 60488.072 & 0 & 0.256 & 15--1000 & COMP & $(1.5 \pm 1.7) \times 10^{-1}$ & -0.55 $\pm$ 0.30 & $(1.1 \pm 0.3) \times 10^{2}$ & \nodata \\
000218 & 58744.596 & 0 & 0.256 & 15--6000 & COMP & $(1.5 \pm 0.4) \times 10^{-1}$ & -0.43 $\pm$ 0.05 & $(9.0 \pm 0.8) \times 10^{2}$ & \nodata \\
 & & 0.256 & 5.888 & 15--6000 & COMP & $(2.3 \pm 0.5) \times 10^{-2}$ & -0.95 $\pm$ 0.06 & $(9.2 \pm 1.7) \times 10^{2}$ & \nodata \\
 & & 5.888 & 13.568 & 15--2000 & PL & $(7.2 \pm 1.9) \times 10^{-3}$ & -1.46 $\pm$ 0.05 & \nodata & \nodata \\
 & & 13.568 & 29.696 & 15--2000 & PL & $(4.8 \pm 1.1) \times 10^{-3}$ & -1.59 $\pm$ 0.05 & \nodata & \nodata \\
000326 & 19134.798 & 0 & 0.256 & 15--1000 & GRB & 1.9 $\pm$ 1.8 & 0.34 $\pm$ 0.61 & $(3.2 \pm 1.0) \times 10^{1}$ & -3.48 $\pm$ 0.52 \\
 & & 0.256 & 8.448 & 15--1000 & GRB & $(7.3 \pm 5.1) \times 10^{-2}$ & -0.31 $\pm$ 0.49 & $(4.8 \pm 1.6) \times 10^{1}$ & -3.50 \\
000420a & 42271.144 & 0 & 0.128 & 15--3000 & COMP & $(4.8 \pm 3.7) \times 10^{-2}$ & -1.03 $\pm$ 0.16 & $(1.9 \pm 1.4) \times 10^{3}$ & \nodata \\
000513 & 40894.793 & 0 & 0.128 & 15--1000 & COMP & $(3.0 \pm 0.5) \times 10^{-2}$ & -1.00 & $(4.7 \pm 1.2) \times 10^{2}$ & \nodata \\
000526 & 84494.896 & 0 & 0.256 & 15--5000 & COMP & $(2.3 \pm 0.6) \times 10^{-1}$ & -0.37 $\pm$ 0.05 & $(4.9 \pm 0.4) \times 10^{2}$ & \nodata \\
000607 & 8689.115 & 0 & 0.064 & 15--5000 & COMP & $(2.2 \pm 1.3) \times 10^{-1}$ & -0.83 $\pm$ 0.13 & $(1.2 \pm 0.4) \times 10^{3}$ & \nodata \\
000608 & 70497.255 & 0 & 0.128 & 15--3000 & COMP & $(8.2 \pm 9.1) \times 10^{-2}$ & -0.91 $\pm$ 0.24 & $(7.0 \pm 3.6) \times 10^{2}$ & \nodata \\
000701b & 25961.013 & 0 & 0.256 & 15--3000 & COMP & $(2.8 \pm 2.3) \times 10^{-2}$ & -0.35 $\pm$ 0.17 & $(7.2 \pm 1.9) \times 10^{2}$ & \nodata \\
 & & 0.256 & 8.448 & 15--1000 & PL & $(4.0 \pm 1.4) \times 10^{-3}$ & -1.18 $\pm$ 0.06 & \nodata & \nodata \\
000727 & 70955.931 & 0 & 0.128 & 15--1000 & COMP & 1.2 $\pm$ 0.6 & -0.55 $\pm$ 0.14 & $(1.0 \pm 0.1) \times 10^{2}$ & \nodata \\
 & & 0.256 & 0.768 & 15--3000 & COMP & $(2.9 \pm 0.7) \times 10^{-1}$ & -1.38 $\pm$ 0.06 & $(3.0 \pm 0.4) \times 10^{2}$ & \nodata \\
 & & 0.768 & 8.704 & 15--3000 & PL & $(1.6 \pm 0.3) \times 10^{-2}$ & -2.30 $\pm$ 0.04 & \nodata & \nodata \\
000818 & 72547.040 & 0 & 0.256 & 15--1000 & PL & $(2.4 \pm 0.9) \times 10^{-2}$ & -1.19 $\pm$ 0.07 & \nodata & \nodata \\
000928 & 6285.374 & 0 & 0.128 & 15--3000 & COMP & $(4.9 \pm 0.6) \times 10^{-2}$ & -0.30 & $(6.1 \pm 0.7) \times 10^{2}$ & \nodata \\
001022 & 20905.666 & 0 & 0.064 & 15--3000 & PL & $(3.2 \pm 1.5) \times 10^{-2}$ & -0.93 $\pm$ 0.07 & \nodata & \nodata \\
001025c & 71369.963 & 0 & 0.192 & 15--5000 & COMP & $(4.7 \pm 3.0) \times 10^{-2}$ & -0.85 $\pm$ 0.13 & $(1.4 \pm 0.6) \times 10^{3}$ & \nodata \\
001204 & 28869.372 & 0 & 0.128 & 15--2000 & PL & $(5.4 \pm 2.4) \times 10^{-2}$ & -1.69$\pm$0.09 & \nodata & \nodata \\
001207b & 34185.588 & 0 & 0.256 & 15--1000 & COMP & $(6.3 \pm 4.0) \times 10^{-2}$ & -1.15 $\pm$ 0.18 & $(2.3 \pm 0.9) \times 10^{2}$ & \nodata \\
 & & 0.256 & 8.448 & 15--500 & COMP & $(3.0 \pm 2.9) \times 10^{-3}$ & -1.62 $\pm$ 0.31 & $(3.9 \pm 6.1) \times 10^{2}$ & \nodata \\
010119 & 37179.556 & 0 & 0.192 & 15--5000 & COMP & $(1.1 \pm 0.3) \times 10^{-1}$ & -0.59 $\pm$ 0.05 & $3.5 \times 10^{2}$ & \nodata \\
010308 & 56338.468 & 0 & 0.256 & 15--2000 & COMP & $(1.8 \pm 0.7) \times 10^{-1}$ & -0.76 $\pm$ 0.10 & $(2.4 \pm 0.4) \times 10^{2}$ & \nodata \\
 & & 0.256 & 8.448 & 15--2000 & COMP & $(7.9 \pm 3.0) \times 10^{-3}$ & -1.11 $\pm$ 0.08 & $2.5 \times 10^{2}$ & \nodata \\
010427 & 67452.969 & 0 & 0.256 & 15--3000 & COMP & $(2.7 \pm 2.1) \times 10^{-2}$ & -0.37 $\pm$ 0.16 & $(8.3 \pm 2.3) \times 10^{2}$ & \nodata \\
010616 & 23724.080 & 0 & 0.064 & 15--1000 & COMP & 1.2 $\pm$ 0.7 & 0.29 $\pm$ 0.16 & $(8.1 \pm 0.8) \times 10^{1}$ & \nodata \\
 & & 0.064 & 0.128 & 15--1000 & COMP & $(8.9 \pm 12.4) \times 10^{-1}$ & 0.06 $\pm$ 0.44 & $(4.5 \pm 1.2) \times 10^{1}$ & \nodata \\
010624 & 48929.130 & 0 & 0.256 & 15--1000 & COMP & $(2.6 \pm 1.3) \times 10^{-1}$ & -0.65 $\pm$ 0.15 & $(9.8 \pm 1.5) \times 10^{1}$ & \nodata \\
010628a & 4206.816 & 0 & 0.256 & 15--5000 & GRB & $(1.5 \pm 0.2) \times 10^{-1}$ & -0.49 $\pm$ 0.20 & $(2.4 \pm 0.7) \times 10^{2}$ & -2.24 $\pm$ 0.20 \\
 & & 0.256 & 8.448 & 15--3000 & COMP & $(7.9 \pm 4.4) \times 10^{-3}$ & -1.15 $\pm$ 0.12 & $(1.3 \pm 0.6) \times 10^{3}$ & \nodata \\
011024 & 74609.296 & 0 & 0.256 & 15--1000 & COMP & $(7.4 \pm 5.9) \times 10^{-2}$ & -1.42 $\pm$ 0.21 & $(5.1 \pm 3.5) \times 10^{2}$ & \nodata \\
 & & 0.256 & 8.448 & 15--500 & PL & $(2.9 \pm 3.8) \times 10^{-3}$ & -2.25 $\pm$ 0.33 & \nodata & \nodata \\
011101 & 34754.534 & 0 & 0.256 & 15--3000 & COMP & $(5.2 \pm 4.7) \times 10^{-2}$ & -0.65 $\pm$ 0.19 & $(6.3 \pm 1.9) \times 10^{2}$ & \nodata \\
020117 & 45909.324 & 0 & 0.256 & 15--1000 & COMP & $(2.0 \pm 1.9) \times 10^{-1}$ & -0.73 $\pm$ 0.29 & $(9.5 \pm 3.2) \times 10^{1}$ & \nodata \\
020306 & 68280.713 & 0 & 0.128 & 15--6000 & COMP & $(4.9 \pm 3.8) \times 10^{-2}$ & -0.47 $\pm$ 0.15 & $(1.1 \pm 0.3) \times 10^{3}$ & \nodata \\
020326 & 39182.941 & 0 & 0.064 & 15--1000 & COMP & $(1.0 \pm 2.1) \times 10^{-1}$ & 0.28 $\pm$ 0.47 & $(1.8 \pm 0.7) \times 10^{2}$ & \nodata \\
020504 & 55835.141 & 0 & 0.256 & 15--5000 & GRB & $(1.3 \pm 0.1) \times 10^{-1}$ & -1.03 $\pm$ 0.10 & $(1.1 \pm 0.4) \times 10^{3}$ & -2.16 $\pm$ 0.39 \\
 & & 0.256 & 8.448 & 15--2000 & PL & $(1.3 \pm 0.0) \times 10^{-2}$ & -1.50 $\pm$ 0.30 & \nodata & \nodata \\
020525a & 16014.630 & 0 & 0.128 & 15--2000 & COMP & $(2.3 \pm 2.6) \times 10^{-2}$ & -0.63 $\pm$ 0.23 & $(1.2 \pm 0.7) \times 10^{3}$ & \nodata \\
020602b & 63030.315 & 0 & 0.256 & 15--1000 & COMP & $(3.0 \pm 2.9) \times 10^{-2}$ & -0.97 $\pm$ 0.23 & $(6.0 \pm 3.7) \times 10^{2}$ & \nodata \\
020715a & 54866.135 & 0 & 0.128 & 15--2000 & COMP & $(7.2 \pm 6.0) \times 10^{-2}$ & -0.56 $\pm$ 0.20 & $(3.2 \pm 0.9) \times 10^{2}$ & \nodata \\
020731a & 1635.905 & 0 & 0.128 & 15--2000 & GRB & $(1.5 \pm 0.1) \times 10^{-1}$ & -0.56 $\pm$ 0.14 & $(3.9 \pm 0.8) \times 10^{2}$ & -3.67 $\pm$ 3.78 \\
020731b & 50231.739 & 0 & 0.128 & 15--1000 & COMP & $(3.9 \pm 9.8) \times 10^{-2}$ & -0.40 $\pm$ 0.63 & $(2.2 \pm 1.8) \times 10^{2}$ & \nodata \\
020828 & 20737.981 & 0 & 0.256 & 15--3000 & COMP & $(3.0 \pm 1.5) \times 10^{-2}$ & -0.98 $\pm$ 0.11 & $(1.6 \pm 0.7) \times 10^{3}$ & \nodata \\
\enddata
\tablenotetext{1}{in designations of section 3}
\tablenotetext{2}{The spectrum of this GRB, accumulated in the time interval 0-0.256~s, was fitted by the sum of
COMP and PL models. The parameters of these models are given in two lines (for COMP and PL model respectively)}.
\end{deluxetable}
\begin{deluxetable}{lcrcrrrlr} 
\tabletypesize{\scriptsize}
\tablecaption{GRBs with afterglow: Fluences and Peak Fluxes\label{Table_tails_basic}}
\tablehead{
\colhead{\parbox[t]{0.9cm}{Burst name}}&%
\colhead{\parbox[t]{1cm}{Figure number}}&%
\colhead{\parbox[t]{0.9cm}{T$_0$, s UT}}&%
\colhead{\parbox[t]{1.5cm}{Energy\\ interval, keV}}&%
\colhead{\parbox[t]{1.5cm}{ Time interval for\\ Fluence, s}} &%
\colhead{\parbox[t]{1.5cm}{Fluence,\\ ergs cm$^{-2}$}} &%
\colhead{\parbox[t]{1.7cm}{Time interval for\\ Peak Flux, s}} &%
\multicolumn{2}{c}{\parbox[t]{4cm}{ \hspace{0.8cm} Peak Flux\\ photons \hspace{0.8cm} ergs \\ cm$^{-2}$s$^{-1}$ \hspace{0.7cm} cm$^{-2}$s$^{-1}$}}
}
\startdata
951014a	& 11 &13108.167 & 15--5000 & -0.088--2.048 &	$2.9 \times 10^{-5}$ & 0.232--0.280 & 	$1.5 \times 10^2$	& $7.3\times 10^{-5}$\\
&13 && 15--3000 & 2.048--50.176 &	$8.2\times 10^{-6}$ & 6.144--10.240 & 1.7 & $3.1\times 10^{-7}$\\
980605 & 66 & 51131.976 & 15--3000 & -0.016--0.140 & $1.4\times 10^{-6}$ & -0.008--0.000 & $1.3\times 10^2$ & $5.1\times 10^{-5}$\\
& 67 &  & 15--1000 & 0.512--70.144 & $3.6\times 10^{-6}$ & 16.896--33.280 & $4.3 \times 10^{-1}$ & $7.6\times 10^{-8}$\\
980706a & 71 & 57586.277 & 15--8000 & -0.056--0.528 & $3.9\times 10^{-5}$ & 0.092--0.108 & $1.9 \times 10^2$ & $1.5\times 10^{-4}$\\
& 72 &  & 15--1000 & 0.768--50.176 & $3.3\times 10^{-6}$ & 2.816--21.504 & $3.0\times 10^{-1}$ & $8.0\times 10^{-8}$\\
981107 & 78 & 781.395 & 15--8000 & -0.008--0.784 & $1.1 \times 10^{-4}$ & 0.340--0.348 & $3.3 \times 10^2$ & $5.2\times 10^{-4}$\\
& 79 &  & 15--1000 & 0.768--90.880 & $6.3\times 10^{-6}$ & 0.768--13.312 & $6.3 \times 10^{-1}$ & $1.6 \times 10^{-7}$\\
990313 & 90 & 33712.652 & 15--2500 & -0.024--0.396 & $1.5 \times 10^{-6}$ & -0.004--0.012 & $9.9 \times 10^1$ & $2.0\times 10^{-5}$\\
& 91 &  & 15--1000 & 0.768--123.648 & $3.1\times 10^{-6}$ & 0.768--61.696 & $4.0 \times 10^{-1}$ & $3.1 \times 10^{-8}$\\
990327 & 92 & 22911.102 & 15--4000 & -0.060--0.116 & $7.0\times 10^{-6}$ & 0.012--0.020 & $1.8 \times 10^2$ & $1.5\times 10^{-4}$\\
& 93 &  & 15--1000 & 0.768--70.400 & $1.2\times 10^{-5}$ & 5.376--8.448 & 4.5 & $1.1\times 10^{-6}$\\
990516 & 97 & 86065.136 & 15--8000 & -0.014--0.432 & $2.4\times 10^{-5}$ & 0.252--0.256 & $2.6 \times 10^2$ & $2.6 \times 10^{-4}$\\
& 99 &  & 15--1000 & 1.024--75.776 & $8.3\times 10^{-6}$ & 10.240--24.576 & 1.6 & $2.3\times 10^{-7}$\\
990712a & 101 & 27915.510 & 15--4000 & -0.080--0.656 & $2.0\times 10^{-5}$ & 0.296--0.312 & $6.1 \times 10^1$ & $5.4\times 10^{-5}$\\
& 102 &  & 15--1000 & 1.024--87.296 & $6.9\times 10^{-6}$ & 27.648--36.096 & 2.5 & $1.9\times 10^{-7}$\\
000218 & 115 & 58744.596 & 15--6000 & -0.176--1.296 & $3.4 \times 10^{-5}$ & 0.104--0.136 & $9.8 \times 10^1$ & $9.2\times 10^{-5}$\\
& 116 &  & 15--1000 & 1.280--63.232 & $2.7\times 10^{-5}$ & 5.120--6.144 & $1.2 \times 10^1$ & $2.7\times 10^{-6}$\\
000701b & 126 & 25961.013 & 15--2500 & -0.176--1.296 & $1.1\times 10^{-5}$ & -0.016--0.016 & $3.3 \times 10^1$ & $2.5\times 10^{-5}$\\
& 127 &  & 15--1000 & 1.280--5.376 & $8.3 \times 10^{-7}$ & 2.816--3.840 & 1.9 & $3.2 \times 10^{-7}$\\
000727 & 129 & 70955.931 & 15--2500 & -0.042--1.040 & $1.1 \times 10^{-5}$ & 0.046--0.056 & $3.9 \times 10^2$ & $5.5\times 10^{-5}$\\
& 131 &  & 15--1000 & 6.704--8.272 & $9.8\times 10^{-6}$ & 6.832--6.960 & $3.0 \times 10^2$ & $2.0\times 10^{-5}$\\
\enddata
\end{deluxetable}
\begin{deluxetable}{lrrrclccc}
\tabletypesize{\scriptsize}
\tablecaption{GRBs with early afterglow: Spectral parameters\label{Table_tails_specpar}}
\tablehead{
\colhead{Burst}&\colhead{T$_0$}& \multicolumn{2}{c}{Time interval} &%
\colhead{Energy} & \colhead{Model\tablenotemark{a}} & \colhead{$\alpha$} & \colhead{$E_0$} & \colhead{$\beta$}\\
\colhead{name}& & \colhead{Start} & \colhead{Stop} &%
\colhead{interval} & & & &\\
& \colhead{s UT} & \colhead{s} & \colhead{s} & \colhead{keV} & & & \colhead{keV} & 
}
\startdata
951014a & 13108.167 & 0 & 0.256 & 15--5000 & GRB & -0.17 $\pm$ 0.13 & $(2.1 \pm 0.3) \times 10^{2}$ & -2.1 $\pm$ 0.1 \\
&  & 0.512 & 7.936 & 15--3000 & GRB &  -1.36 $\pm$ 0.12& $(4.5 \pm 1.8) \times 10^{2}$ & -2.2 $\pm$ 0.1\\
980605 & 51131.976 & 0 & 0.064 & 15--5000 & COMP & -0.74 $\pm$ 0.25 & $(1.0 \pm 0.8) \times 10^{3}$ & \nodata \\
&  & 0.512 & 70.144 & 15--1000 & COMP &  -1.04 $\pm$ 0.40 & $(2.3 \pm 1.8) \times 10^{2}$ & \nodata\\
980706a & 57586.277 & 0 & 0.256 & 15--6000 & COMP & -0.75 $\pm$ 0.03 & $(1.3 \pm 0.1) \times 10^{3}$ & \nodata \\
&  & 0.768 & 50.176 & 15--1000 & COMP & -0.94 $\pm$ 0.45 & $(3.6 \pm 3.3) \times 10^{2}$ & \nodata\\
981107 & 781.395 & 0 & 0.128 & 15--8000 & COMP & -0.08 $\pm$ 0.16 & $(8.0 \pm 1.3) \times 10^{2}$ & \nodata \\
&  & 0.768 & 90.88 & 15--1000 & PL & -1.44 $\pm$ 0.13 & \nodata & \nodata\\
990313 & 33712.652 & 0 & 0.128 & 15--2000 & GRB & -1.08 $\pm$ 0.18 & $(2.0 \pm 0.9) \times 10^{2}$ & -2.4 $\pm$ 0.4 \\
&  & 0.768 & 123.648 & 15--1000 & PL & -2.03 $\pm$ 0.17 & \nodata & \nodata\\
990327 & 22911.102 & 0 & 0.064 & 15--5000 & COMP & -0.86 $\pm$ 0.11 & $(2.1 \pm 0.8) \times 10^{3}$ & \nodata \\
&  & 0.768 & 70.4 & 15--1000 & COMP & -1.07 $\pm$ 0.20 & $(3.3 \pm 1.2) \times 10^2$ & \nodata\\
990516 & 86065.136 & 0 & 0.064 & 15--5000 & COMP & -1.26 $\pm$ 0.16 & $(1.9 \pm 2.4) \times 10^{3}$ & \nodata \\
&  & 1.024 & 75.776 & 15--1000 & PL & -1.81 $\pm$ 0.09 & \nodata & \nodata\\
990712a & 27915.510 & 0 & 0.256 & 15--5000 & COMP & -0.20 $\pm$ 0.13 & $(6.1 \pm 1.0) \times 10^{2}$ & \nodata \\
&  & 1.024 & 96.768 & 15--1000 & PL & -2.31 $\pm$ 0.14 & \nodata & \nodata\\
000218 & 58744.596 & 0 & 0.256 & 15--6000 & COMP & -0.43 $\pm$ 0.05 & $(9.0 \pm 0.8) \times 10^{2}$ & \nodata \\
&  & 1.28 & 70.144 & 15--1000 & PL & -1.58 $\pm$ 0.30 & \nodata & \nodata\\
000701b & 25961.013 & 0 & 0.256 & 15--3000 & COMP & -0.35 $\pm$ 0.17 & $(7.2 \pm 1.9) \times 10^{2}$ & \nodata \\
&  & 1.28 & 5.376 & 15--1000 & COMP & -1.44 $\pm$ 0.30 & $(5.6 \pm 3.7) \times 10^2$  & \nodata\\
000727 & 70955.931 & 0 & 0.128 & 15--1000 & COMP & -0.55 $\pm$ 0.14 & $(1.0 \pm 0.1) \times 10^{2}$ & \nodata \\
&  & 0.768 & 8.704 & 15--2500 & PL & -2.50 $\pm$ 0.04 & \nodata & \nodata \\
\enddata
\end{deluxetable}
\end{document}